\documentclass[conference]{IEEEtran}
\usepackage{subfigure}
\usepackage{algorithm}
\usepackage{algorithmic}
\usepackage{color}
\usepackage{stmaryrd}
\usepackage{amsfonts}
\usepackage{amsmath}
\usepackage{amssymb}
\usepackage{amsthm,empheq}
\usepackage{cite}
\usepackage{slashbox}
\usepackage[dvips]{graphicx}
\usepackage{hyperref}
\usepackage{enumerate}
\usepackage{mathrsfs}

\newtheorem{definition}{Definition}
\newtheorem{lemma}{Lemma}
\newtheorem{corollary}{Corollary}
\newtheorem{remark}{Remark}
\newtheorem{proposition}{Proposition}

\begin{document}
\title{Proof of Control of a UAV and a UGV Cooperating to Manipulate an Object}

\author{%
Tam Nguyen, Emanuele Garone\\[3mm]\\[3mm]
}
\maketitle
\begin{abstract}
This paper focuses on the control of a system composed of an Unmanned Aerial Vehicle (UAV) and an Unmanned Ground Vehicle (UGV) which cooperate to manipulate an object. The two units are subject to actuator saturations and cooperate to move the object to a desired pose, characterized by its position and inclination. The paper proposes a control strategy where the ground vehicle is tasked to deploy the object to a certain position, whereas the aerial vehicle adjusts its inclination. The ground vehicle is governed by a saturated proportional-derivative control law. The aerial vehicle is regulated by means of a cascade control specifically designed for this problem that is able to exploit the mechanical interconnection. The stability of the overall system is proved through Input-to-State Stability and Small Gain theorem arguments. To solve the problem of constraints satisfaction, a nonlinear Reference Governor scheme is implemented. Numerical simulations are provided to demonstrate the effectiveness of the proposed method.
\end{abstract}

\section{Introduction}

The use of Unmanned Aerial Vehicles (UAVs) as aerial manipulators has recently drawn the attention of several researchers around the world \cite{pounds2011yale,nicotra2014nested,maza2010multi,mellinger2013cooperative,gioioso2014flying,michael2011cooperative,korpela2012mm,lippiello2012exploiting,arleo2013control}. Early experiments conducted in controlled lab environments have demonstrated the transportation (control of the position) \cite{pounds2011yale,nicotra2014nested,maza2010multi,mellinger2013cooperative} and manipulation (control of the position and orientation) \cite{gioioso2014flying,korpela2012mm,michael2011cooperative,lippiello2012exploiting,arleo2013control} of objects through UAVs.

Most of the works on this subject concern the transportation of objects, through single and multiple UAVs, including grasping \cite{pounds2011yale}, hovering capture, load stability \cite{nicotra2014nested}, and cooperative transportation \cite{maza2010multi,mellinger2013cooperative}. For what concerns the manipulation of objects through UAVs, only a few preliminary works have been proposed. These works include the manipulation of objects through a team of UAVs \cite{gioioso2014flying} or through a single UAV equipped with robotic arms \cite{korpela2012mm}. In \cite{michael2011cooperative}, a triangular object suspended by cables is manipulated through three UAVs.

The physical interaction between UAVs and Unmanned Ground Vehicles (UGVs) has recently attracted some interest and represents a relatively young research topic. Early works on the subject include the pulling of a cart through one or two quadrotors \cite{srikanth2011controlled}, the cooperative pose stabilization of a UAV through a team of ground robots \cite{naldi2012cooperative}, and the modeling \cite{muttin2011umbilical} and control \cite{nicotra2014taut,tognon2015nonlinear} of tethered UAVs.

This paper proposes a control framework for the manipulation of an object through a heterogeneous team consisting of a UAV cooperating with a UGV. Both vehicles are subject to actuator saturations. The idea of manipulating objects using a team of autonomous aerial and ground vehicles is, at the best of the authors' knowledge, new, and has potential applications in the world of Autonomous Robotic Construction (ARC) \cite{willmann2012aerial,augugliaro2014flight,augugliaro2013building,mirjan2013architectural,lindsey2011construction} as it could allow UAVs to collaborate with ground vehicles to build structures in human-denied environments by helping them positioning beams and bars.

The paper is organized as follows. First, the equations of motion of the UGV-object-UAV system are derived using the Euler-Lagrange approach. Then, the attainable configurations of equilibrium of the system are discussed taking into account the saturations of the actuators. Afterwards, a decentralized control architecture is proposed, where the stability of the system is proved through Input-to-State Stability and Small Gain arguments. In order to ensure constraints satisfaction, the control law is augmented with a nonlinear Reference Governor \cite{bemporad1998reference,kolmanovsky2014reference}. Numerical simulations are provided to demonstrate the effectiveness of the proposed solution.

\section{Notations}

\begin{definition}\label{saturation}
The saturation function $\sigma_\lambda$ is defined as
\begin{equation}
\sigma_\lambda(x):=\text{sign}(x)\text{min}{(|x|,\lambda)},
\end{equation}
where $\lambda\in\mathbb{R}_0^+$.
\end{definition}

\begin{definition}\label{strictpositivesaturation}
The positive saturation function $\sigma_{0,\lambda}$ is defined as
\begin{equation}
\sigma_{0,\lambda}(x):=
\begin{aligned}
\begin{cases}
\sigma_{\lambda}(x), &\mbox{if $x\geq0$}\\
0, &\mbox{if $x<0$},
\end{cases}
\end{aligned}
\end{equation}
where $\lambda\in\mathbb{R}_0^+$.
\end{definition}

\section{Problem Statement}

\begin{figure}[!h]
\includegraphics[scale=0.3]{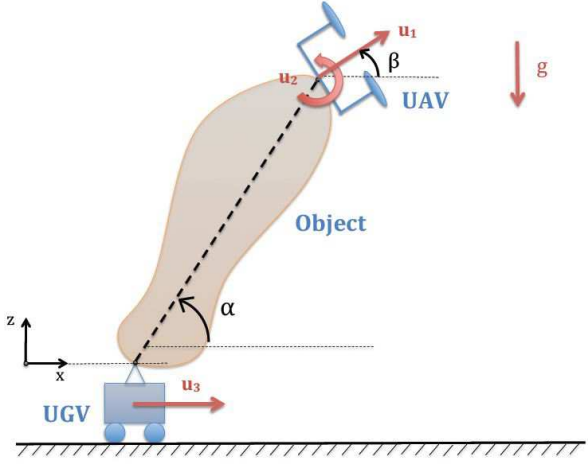}
\caption{Planar model of a UAV and a UGV manipulating a rigid body.}\label{model}
\end{figure}

Consider the planar model of a UGV and of a quadrotor UAV manipulating a rigid body as depicted in Fig. \ref{model}. It is assumed that the joints between the bodies are ideal and the center of mass of the UAV coincides with the joint position.

The UAV has mass $m_u\in\mathbb{R}_{0}^+$ and moment of inertia $\mathcal{I}_u\in\mathbb{R}_0^+$. The UGV has mass $m_c\in\mathbb{R}_0^+$. The object has mass $m_b\in\mathbb{R}_0^+$, moment of inertia $\mathcal{I}_b\in\mathbb{R}_0^+$, and length $L\in\mathbb{R}_0^+$. Its center of mass is at distance $d_G\in\mathbb{R}_0^+$ from the UGV. Let the position of the cart $x\in\mathbb{R}$, the inclination of the object $\alpha\in[0,\pi]$, and the attitude of the UAV $\beta\in[0,2\pi]$ be the generalized coordinates of the system. All the angles are defined with respect to the horizon.

The bodies are subject to the gravity acceleration $g$. The UAV propellers generate a total thrust $u_1\in\mathbb{R}^+$ and a resultant torque $u_2\in\mathbb{R}$. The UGV motors produce a force $u_3\in\mathbb{R}$. The signs of $u_1,u_2$, and $u_3$ are defined positive with respect to the oriented vectors depicted in Fig. \ref{model}. The saturations of $u_1,u_2$, and $u_3$ are
\begin{equation}\label{actuatorsconstraints}
\begin{cases}
\begin{aligned}
0&\leq u_1\leq&U_{max}\\
-T_{max}&\leq u_2 \leq &T_{max}\\
-F_{max}& \leq u_3 \leq& F_{max},
\end{aligned}
\end{cases}
\end{equation}
where $U_{max},T_{max},F_{max}\in\mathbb{R}_0^+$.

To derive the equations of motion, the Euler-Lagrange method is used. To this end, consider the kinetic and potential energies of the system $\mathcal{T}$ and $\mathcal{V}$, respectively, which are
\begin{eqnarray*}
\begin{cases}
\mathcal{T}=&\frac{1}{2}m_c\dot{x}^2+\frac{1}{2}m_b(\dot{x}^2-2\dot{x}d_G\dot{\alpha}\sin{\alpha}+d_G^2\dot{\alpha}^2)\\&+\frac{1}{2}m_u(\dot{x}^2-2\dot{x}L\dot{\alpha}\sin{\alpha}+L^2\dot{\alpha}^2)+\frac{1}{2}\mathcal{I}_{b}\dot{\alpha}^2+\frac{1}{2}\mathcal{I}_u\dot{\beta}^2\\
\mathcal{V}=&m_bd_G\sin{\alpha}g+m_{u}L\sin{\alpha}g.
\end{cases}
\end{eqnarray*}
Assuming friction forces negligible, the principle of the least action $\frac{d}{dt}\frac{\partial\mathcal{L}}{\partial{\dot{q_i}}}-\frac{\partial{\mathcal{L}}}{\partial{q_i}}=f_i$ can be used, where $\mathcal{L}=\mathcal{T}-\mathcal{V}$ is the Langrangian, $q_i$ are the generalized coordinates, and $f_i$ the external forces acting on the system. The equations of motion of the system are
\begin{align}\label{complex}
\begin{cases}
M_{tot}\ddot{x}-ML(\sin{\alpha}\ddot{\alpha}+\dot{\alpha}^2\cos{\alpha})&=u_3\\
M(-\sin{\alpha}\ddot{x}+\cos{\alpha}g)+\mathcal{I}_0\ddot{\alpha}&=u_1\sin(\beta-\alpha)\\
\mathcal{I}_u\ddot{\beta}&=u_2,
\end{cases}
\end{align}
where $M_{tot}=m_c+m_b+m_u$ is the total mass of the system, $M=\frac{m_bd_G}{L}+m_u$ the apparent mass of the UAV and the object, and $\mathcal{I}_0=\frac{m_bd_G^2+\mathcal{I}_b}{L}+m_uL$ the moment of inertia of the system divided by the length of the object $L$. For the sake of simplicity, it is assumed that $m_c>>m_b$ and $m_c>>m_u$ so that the effect of the UAV dynamics on the UGV can be neglected. Consequently, the system becomes
\begin{eqnarray}
\begin{aligned}
\begin{cases}
\label{UGV}m_c\ddot{x}&=u_3\\
\label{bar}M(-\sin{\alpha}\ddot{x}+\cos{\alpha}g)+\mathcal{I}_0\ddot{\alpha}&=u_1\sin{(\beta-\alpha)}\\
\label{UAV}\mathcal{I}_u\ddot{\beta}&=u_2.
\end{cases}
\end{aligned}
\end{eqnarray}

The objective of this paper is to control the pose of the object to a desired angle $\bar{\alpha}$ and position $\bar{x}$. To this end, the first step is to analyze the attainable configurations of equilibrium considering the saturations of the actuators.

\section{Attainable Configurations of Equilibrium}

In this section, the attainable configurations of equilibrium $[\bar{x},\bar{\alpha},\bar{\beta}]^T$ and the associated steady state input vector $\bar{u}=[\bar{u}_1,\bar{u}_2,\bar{u}_3]^T$ are discussed taking into account the saturations of the UAV. Setting all the time derivatives of (\ref{UGV}) to zero, it follows that the configurations of equilibrium must satisfy the system of equations
\begin{eqnarray}
\begin{aligned}
\begin{cases}
\label{u2eq} \bar{u}_2&=0\\ \label{u3eq}
\bar{u}_3&=0\\
Mg\cos{\bar{\alpha}}&=\bar{u}_1\sin{(\bar{\beta}-\bar{\alpha})}. \label{u1eq}
\end{cases}
\end{aligned}
\end{eqnarray}
Clearly, the first two equations of (\ref{u2eq}) give $\bar{u}_2=0$ and $\bar{u}_3=0$ as the only input associated to an equilibrium. Moreover, any $\bar{x}\in\mathbb{R}$ is an attainable point of equilibrium since $\bar{x}$ does not appear in (\ref{u3eq}). For what concerns the last equation of (\ref{u1eq}), due to the saturation (\ref{actuatorsconstraints}) of $\bar{u}_1$, the magnitude of $\bar{u}_1\sin(\bar{\beta}-\bar{\alpha})$ is maximal when
\begin{align}\label{maxthrust}
\begin{cases}
\bar{u}_1&=U_{max}\\
\bar{\beta}&=\bar{\alpha}\pm\pi/2.
\end{cases}
\end{align}
Accordingly, there are two possible cases.
\begin{enumerate}
\item If $U_{max}\geq Mg$, any $\bar{\alpha}\in[0,\pi]$ is an attainable angle of equilibrium for the object.
\item If $U_{max}<Mg$, the attainable angles of equilibrium are restricted to the interval $\bar{\alpha} \in [\alpha_{min}, \alpha_{max}]$, the boundaries of which are
\begin{align}\label{constraintsalpha}
\begin{cases}\alpha_{min}=&\arccos{\frac{U_{max}}{Mg}} \\ \alpha_{max}=&\arccos{\frac{-U_{max}}{Mg}}.\end{cases}
\end{align}
\end{enumerate}

Finally note that, for a given steady-state angle $\bar{\alpha}$, the attainable equilibria for the attitude $\bar{\beta}$ are restricted to the interval $\bar{\beta} \in [\beta_{min},\beta_{max}]$. The boundaries of this interval can be computed solving (\ref{u1eq}) with $\bar{u}_1=U_{max}$ and are
\begin{equation}\label{constraintsbeta}
\begin{cases}
\begin{aligned}
\beta_{min}&=\begin{cases}
\arcsin{\left(\frac{Mg\cos{\bar{\alpha}}}{U_{max}}\right)}+\bar{\alpha},\text{if $\alpha_{min}\leq \bar{\alpha} < \pi/2$}\\
\arcsin{\left(\frac{-Mg\cos{\bar{\alpha}}}{U_{max}}\right)}+\bar{\alpha},\text{if $\pi/2\leq \bar{\alpha} \leq \alpha_{max}$}
\end{cases}\\
\beta_{max}&=\begin{cases}
\pi-\arcsin{\left(\frac{Mg\cos{\bar{\alpha}}}{U_{max}}\right)}+\bar{\alpha},\text{if $\alpha_{min}\leq \bar{\alpha} < \pi/2$}\\
\pi-\arcsin{\left(\frac{-Mg\cos{\bar{\alpha}}}{U_{max}}\right)}+\bar{\alpha},\text{if $\pi/2\leq \bar{\alpha} \leq \alpha_{max}$}.
\end{cases}
\end{aligned}
\end{cases}
\end{equation}

\section{Control Architecture}

The proposed control architecture consists of two separate control units in charge of governing the UGV and the UAV. The UGV control loop generates a control input $u_3$ such that $x(t)$ asymptotically tends to $\bar{x}$. The UAV control loop is tasked to regulate the inclination of the transported object. The proposed UAV controller uses a cascade control approach, where the inner loop controls the UAV attitude and the outer loop controls the inclination of the object. The overall asymptotic stability of the system is proved assuming a Proportional-Derivative (PD) controller for the UAV attitude. 

\begin{figure}
\centering{
\includegraphics[scale=0.25]{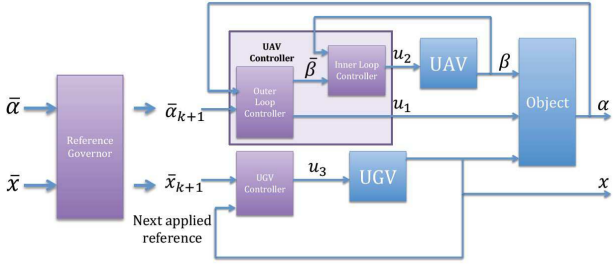}
\caption{Decentralized control architecture.}\label{controlarchitecture}}
\end{figure}

For constraints satisfaction, a nonlinear Reference Governor (RG) is added to the scheme. Whenever necessary, the RG modifies the references to ensure the non-violation of the constraints. The complete control architecture is depicted in Fig. \ref{controlarchitecture}. The design of the controllers are detailed in the next sections.

%
%
\section{UGV Control}\label{UGVcontrolloop}
The objective of the UGV control loop is to steer the UGV to a desired position $\bar{x}$. To this end, the nested saturated PD control law
\begin{equation}\label{u3}
u_3=-\sigma_{\lambda_1}(k_{d,x}\dot{x}+\sigma_{\lambda_2}(k_{p,x}(x-\bar{x})))
\end{equation}
is proposed, where $k_{p,x},k_{d,x}\in\mathbb{R}_0^+$ are the parameters to be tuned and $\sigma_{\lambda_1},\sigma_{\lambda_2}$ are the saturation functions (cf. Definition \ref{saturation}). The choice of $\lambda_1\leq F_{max}$ ensures the satisfaction of the saturation constraint on $u_3$. The following Lemma summarizes the main properties of this control law.
\begin{lemma}
Consider the UGV in (\ref{UGV}) controlled by the saturated PD (\ref{u3}).  
\begin{enumerate}[i]\label{Lemma}
\item The closed loop system is Globally Asymptotically Stable (GAS) for any desired point of equilibrium $\bar{x}$ and for any $k_{p,x}>0$, $k_{d,x}>0$, and $\lambda_2 < \dfrac{1}{2} \lambda_1 k_{d,x}.$
\item The acceleration $\ddot{x}$ is bounded by $\lambda_1/m_c$ and, for a constant desired point of equilibrium $\bar{x}$, vanishing in time, i.e. $lim_{t \rightarrow \infty} \ddot{x}=0$.\label{lemma}
\end{enumerate}
\begin{proof}
The proof can be found in \emph{Lemma 1} of \cite{marconi2000robust}.
\end{proof}
\end{lemma}
%
%

\section{UAV Control}\label{UAVcontrolloop}
The objective of the UAV control loop is to ensure that $\lim_{t\to\infty}\alpha(t)=\bar{\alpha}$. To this end, a cascade strategy approach is proposed, where the outer loop controller (see Fig. \ref{controlarchitecture}) is firstly designed, assuming the inner loop ideal. Then, the stability of the system is proved using a PD for the inner control loop.

\subsection{Ideal Attitude Dynamics}
Given a desired UAV attitude $\bar{\beta}$, assume for the moment that the attitude dynamics is ideal, and therefore $\beta(t)=\bar{\beta}$ at each instant $t$. As a consequence, the second equation of (\ref{bar}) becomes
\begin{equation}\label{u1ex}
\mathcal{I}_0\ddot{\alpha}=u_1\sin\bar{\theta}-Mg\cos{\alpha}+d,
\end{equation}
where $\bar{\theta}:=\bar{\beta}-\alpha$ is the desired relative attitude of the UAV and $d:=-M\ddot{x}\sin{\alpha}$ the external disturbance induced by the UGV. Define $f_t$ the tangential force induced by the UAV
\begin{equation}\label{ft}
f_t:=u_1\sin{\bar{\theta}}.
\end{equation}
Eq. (\ref{u1ex}) becomes
\begin{equation}\label{u1imp}
\mathcal{I}_0\ddot{\alpha}=f_t-Mg\cos{\alpha}+d.
\end{equation}
$f_t$ can be used as an input to control the dynamics of $\alpha$. The proposed control law is a PD with gravity compensation
\begin{equation}\label{ftexplicit}
f_t=-k_{p,\alpha}(\alpha-\bar{\alpha})-k_{d,\alpha}\dot{\alpha}+Mg\cos{\alpha},
\end{equation}
where $k_{p,\alpha},k_{d,\alpha}\in\mathbb{R}_0^+$ are the parameters to be tuned. Eq. (\ref{u1imp}) controlled by (\ref{ftexplicit}) is $$\mathcal{I}_0\ddot{\alpha}=-k_{p,\alpha}(\alpha-\bar{\alpha})-k_{d,\alpha}\dot{\alpha}+d,$$which is a linear system and is therefore Input-to-State Stable (ISS) with respect to the disturbance $d$ for any $k_{p,\alpha}>0$ and $k_{d,\alpha}>0$. Since $d$ is bounded and asymptotically tending to zero (Lemma \ref{Lemma}), $\lim_{t\to\infty}\alpha(t)=\bar{\alpha}$ in absence of attitude dynamics.

At this point, it remains to determine a couple $u_1$ and $\bar{\theta}$ which produces the tangential force $f_t$. In line of principle, Eq. (\ref{ft}) admits an infinite number of solutions. Rewriting (\ref{ft}) in the form
\begin{equation}\label{mapu1}
u_1=\dfrac{f_t}{\sin{\bar{\theta}}},
\end{equation}
the following continuous mapping is proposed in this paper
\begin{equation}\label{maptheta}
\bar{\theta}=\sigma_{\pi/2}(\gamma\arctan{(\epsilon f_t)}),
\end{equation}
where $\sigma_{\pi/2}$ is the saturation function limiting the variable to $\pm\pi/2$, and $\gamma,\epsilon\in\mathbb{R}_0^+$ are parameters to be chosen such that thrust constraints are satisfied. Note that this mapping always guarantees the positiveness of $u_1$. In fact, both $f_t$ and $\sin(\sigma_{\pi/2}(\gamma\arctan{(\epsilon f_t)}))$ are odd and monotonically increasing functions. In view of (\ref{mapu1}), the quotient of two odd and monotonically increasing functions is always positive. Remark also that, with the mapping (\ref{maptheta}), $u_1$ does not present any singularities since $\lim_{f_t\to0}u_1=\dfrac{1}{\gamma\epsilon}$.

To choose the parameter $\gamma$, the saturation (\ref{actuatorsconstraints}) on $u_1$ must be satisfied when $f_t=U_{max}$. It follows from (\ref{mapu1}) that, in this case, $\bar{\theta}$ must be equal to $\pi/2$. As a result, following from (\ref{maptheta}), $\gamma$ must satisfy
\begin{equation}\label{gamma}
\gamma = \dfrac{\pi}{2\arctan(\epsilon U_{max})}.
\end{equation}
For what concerns the choice of $\epsilon$, as clarified in the following Lemma, steady-state constraints are always ensured for any $\epsilon\in\mathbb{R}^+_0$ and therefore, $\epsilon$ can be freely chosen as a tuning parameter.

\begin{lemma}\label{epsilontuning}
For any $\epsilon\in\mathbb{R}_0^+$, the mapping (\ref{maptheta}) with $\gamma$ satisfying (\ref{gamma}) ensures $|\bar{u}_1|\leq U_{max}$.
\begin{proof}
Consider first $f_t\in[0,U_{max}]$. In view of (\ref{ftexplicit}), the control input $f_t$ at equilibrium must be $$f_t=Mg\cos\bar{\alpha},$$where $\bar{\alpha}\in[0,\pi/2]$. Define the minimum relative UAV attitude $\bar{\theta}_{min}:=\beta_{min}-\bar{\alpha}$. Following from (\ref{constraintsbeta}), $\bar{\theta}_{min}$ is 
$$\bar{\theta}_{min}=\arcsin\left(\frac{f_t}{U_{max}}\right).$$
Because of the third equation of (\ref{u3eq}), to ensure $|\bar{u_1}|\leq U_{max}$ for all points of equilibrium, the inequality $\bar{\theta}\geq\bar{\theta}_{min}$ must be satisfied for $f_t\in[0,U_{max}]$. Choosing $\gamma$ as in (\ref{gamma}), the inequality $$\gamma\arctan{(\epsilon f_t)}\geq\arcsin{(f_t/U_{max})}$$ holds true for $\epsilon\in\mathbb{R}_0^+$ since, if restricted to $f_t\in[0,U_{max}]$, $\gamma\arctan{(\epsilon f_t)}$ is convex and $\arcsin(f_t/U_{max})$ is concave (see Fig. \ref{atan}).
The same arguments hold true for $f_t\in[-U_{max},0]$, where $\bar{\theta}\leq\bar{\theta}_{max}$ with $\bar{\theta}_{max}:=\beta_{max}-\bar{\alpha}$, concluding the proof.
\begin{figure}[!h]\centering{
\includegraphics[scale=0.55]{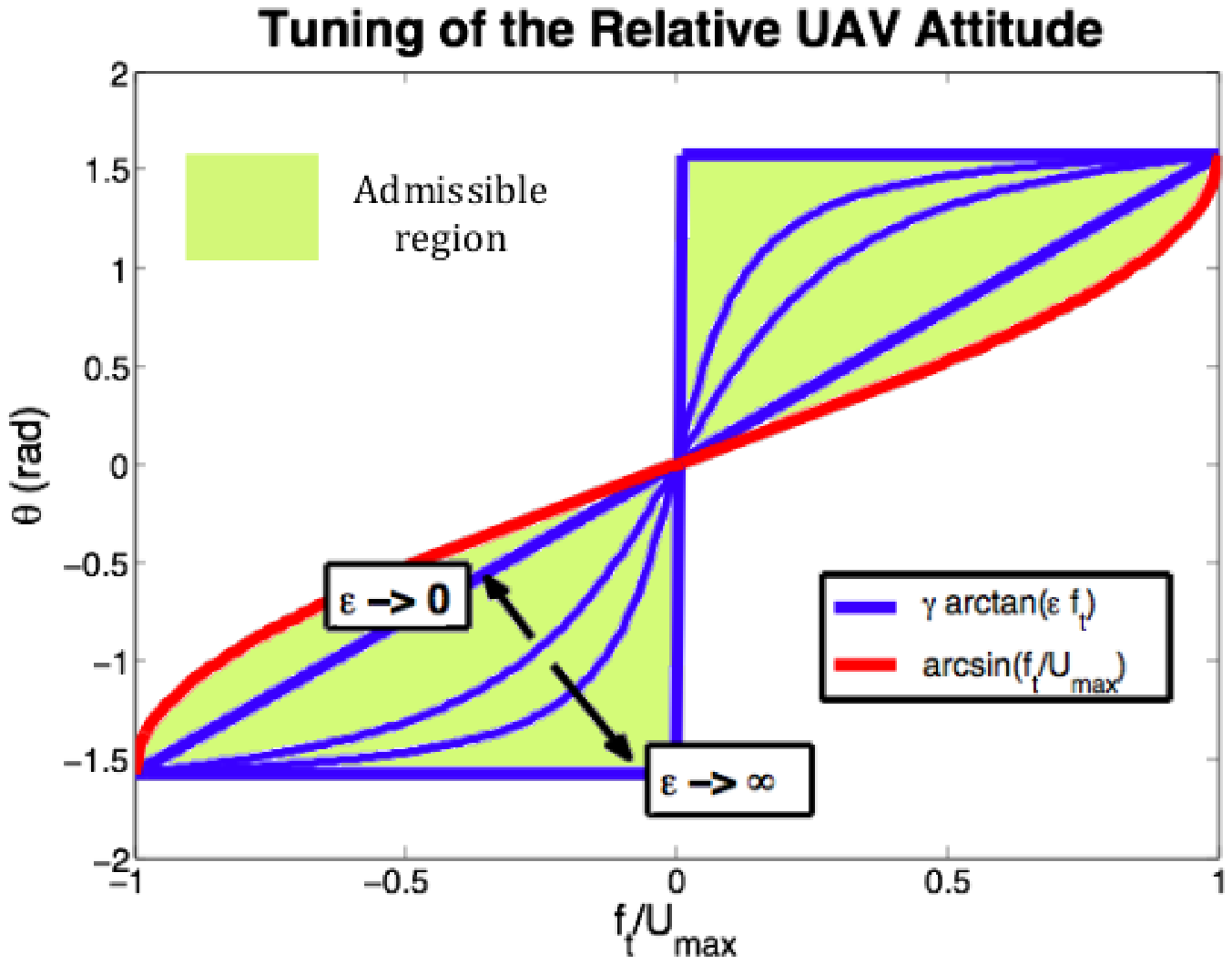}
\caption{Attitude reference $\bar{\theta}$ tuned with $\epsilon\to0$ and $\epsilon\to\infty$.}\label{atan}}
\end{figure}
\end{proof}
\end{lemma}
\begin{remark}
For design purposes, the mapping of $\bar{\theta}$ can be modified with $\epsilon$ to tune the response of the UAV with respect to a change of $f_t$.
\end{remark}

\subsection{Presence of Attitude Dynamics}

In this subsection, the system dynamics seen in the previous subsection is analyzed in the case of a non-ideal attitude dynamics controlled by a PD. Define the attitude error $\tilde{\beta}:=\beta-\bar{\beta}.$ Clearly, the error on the attitude is equal to the error on $\theta$
\begin{equation}\label{thetaequalbeta}
\tilde{\theta} =\tilde{\beta},
\end{equation}
where $\tilde{\theta}:=\theta-\bar{\theta}$. Therefore, in the presence of an attitude error, Eq. (\ref{u1ex}) becomes
\begin{equation}\label{prova}
\mathcal{I}_0\ddot{\alpha}=u_1\sin{(\bar{\theta}+\tilde{\theta})}-Mg\cos{\alpha}+d.
\end{equation}
Developing the sine term, (\ref{prova}) becomes
\begin{equation}\label{provb}
\mathcal{I}_0\ddot{\alpha}=u_1\sin\bar{\theta}\cos\tilde{\theta}+u_1\cos{\bar{\theta}}\sin{\tilde{\theta}}-Mg\cos\alpha+d.
\end{equation}
Expressing $u_1$ by the mapping (\ref{mapu1}), Eq.(\ref{provb}) becomes
\begin{equation}\label{provc}
\mathcal{I}_0\ddot{\alpha}=f_t\cos\tilde{\theta}+u_1\cos{\bar{\theta}}\sin{\tilde{\theta}}-Mg\cos\alpha+d.
\end{equation}
Using the control law (\ref{ftexplicit}) in (\ref{provc}), it follows that
\begin{equation}\label{u1experrorcanonic}
\mathcal{I}_0\ddot{\alpha}=(-k_{p,\alpha}(\alpha-\bar{\alpha})-k_{d,\alpha}\dot{\alpha})\cos{\tilde{\theta}}+{\delta_{\tilde{\theta}}}+d,
\end{equation}
where
\begin{equation}\label{deltatildeequation}
{\delta_{\tilde{\theta}}}:=u_1\cos{\bar{\theta}}\sin{\tilde{\theta}}-Mg\cos\alpha(1-\cos\tilde{\theta}),
\end{equation}
which is the disturbance induced by the attitude error $\tilde{\theta}$. The following Lemma states that this disturbance is bounded for any $f_t\in\mathbb{R}$.
\begin{lemma}\label{u1coslemma}
The disturbance $\delta_{\tilde{\theta}}$ is bounded and satisfies $|\delta_{\tilde{\theta}}|\leq(2/\pi|\sin\tilde{\theta}|+Mg|1-\cos\tilde{\theta}|)$ for any $\epsilon\in\mathbb{R}_0^+$ and any $f_t\in\mathbb{R}$.
\begin{proof}
The proof can be found in Appendix \ref{u1coslemmaproof}.
\end{proof}
\end{lemma}
\begin{remark}
It is clear that the disturbance $\delta_{\tilde{\theta}}$ vanishes when $\tilde{\theta}\to0$.
\end{remark}

Eq. (\ref{u1experrorcanonic}) is the outer loop in the presence of an attitude error, where the states $[\alpha,\dot{\alpha}]^T$ are affected by the exogenous inputs $\delta_{\tilde{\theta}}$ and $d$. For this system, the following result holds true.
\begin{proposition}\label{propositionLyapunov}
Consider the outer loop (\ref{u1experrorcanonic}) for any $k_{p,\alpha}>0$ and $k_{d,\alpha}>0$ and for $\tilde{\theta}\in[-\tilde{\theta}_{max},\tilde{\theta}_{max}]$ where $\tilde{\theta}_{max}\in(-\pi/2,\pi/2)$.
\begin{enumerate}[i]
\item The system is ISS with restriction $\tilde{\theta}\in[-\tilde{\theta}_{max},\tilde{\theta}_{max}]$ with respect to $d$.
\item The system is ISS with restriction $\tilde{\theta}\in[-\tilde{\theta}_{max},\tilde{\theta}_{max}]$ with respect to $\tilde{\beta}$.
\item The asymptotic gain $\gamma_{out}$ between $\tilde{\beta}$ and $\dot{\bar{\beta}}$ is finite.
\end{enumerate}
\begin{proof}
The proof can be found in Appendix \ref{proofLyapunov}.
\end{proof}
\end{proposition}

At this point, consider the inner attitude dynamics described by the third equation of (\ref{UAV}). To control the inner loop, a PD control law is chosen:
\begin{equation}\label{u2control}
u_2=-k_{p,\beta}\tilde{\beta}-k_{d,\beta}\dot{\beta},
\end{equation}
where $k_{p,\beta},k_{d,\beta}\in\mathbb{R}_0^+$ are control parameters to be tuned. The attitude error dynamics $\dot{\tilde{\beta}}$ becomes
\begin{align}\label{attitudedynamics}
\begin{cases}
\dot{\tilde{\beta}}&=\dot{\beta}-\dot{\bar{\beta}}\\
\mathcal{I}_u\ddot{\beta}&=-k_{p,\beta}\tilde{\beta}-k_{d,\beta}\dot{\beta}.
\end{cases}
\end{align}
System (\ref{attitudedynamics}) is the inner loop, where the states $[\tilde{\beta},\dot{\beta}]^T$ are affected by the exogenous input $\dot{\bar{\beta}}$. The following result can be proved.
%
%
\begin{proposition}\label{UAVAttitudeProposition}
The inner loop system (\ref{attitudedynamics}) is ISS with respect to $\dot{\bar{\beta}}$ for any $k_{p,\beta}>0$ and $k_{d,\beta}>0$. The asymptotic gain $\gamma_{in}$ between the disturbance $\dot{\bar{\beta}}$ and the output $\tilde{\beta}$ is finite and can be made arbitrarily small for sufficiently large $k_{p,\beta}>0$ and $k_{d,\beta}>0.$

\begin{proof}
The proof can be found in Appendix \ref{proofUAVAttitude}.
\end{proof}
\end{proposition}

Using ISS and Small Gain arguments, it is possible to prove the asymptotic stability of the overall system.
\begin{proposition}\label{smallgainproposition}
Consider (\ref{UGV}) controlled by (\ref{u3}), (\ref{ftexplicit})-(\ref{maptheta}) and (\ref{u2control}). Given a desired position $\bar{x} \in \mathbb{R},$  a desired inclination $\bar{\alpha} \in [\alpha_{min}, \alpha_{max}],$ and the resulting steady-state attitude $\bar{\beta}$, the point of equilibrium $[\bar{x}, \bar{\alpha}, \bar{\beta}]^T$  is asymptotically stable for suitably large $k_{p,\beta}$ and $k_{d,\beta}$ in absence of the saturations (\ref{actuatorsconstraints}) for any initial condition satisfying
\begin{equation}\label{smallgaininitialcondition}
\sqrt{\theta^2(0)+\dot{\theta}^2(0)}+\gamma_{in}(\sqrt{\alpha^2(0)+\dot{\alpha}^2(0)})<(1-\gamma_{in}\gamma_{out})|\tilde{\theta}_{max}|.
\end{equation}

\begin{proof}
From Propositions 1 and 2, $\gamma_{in}$ and $\gamma_{out}$ are proven to be finite. Since $\gamma_{in}$ can be made arbitrarily small with sufficiently large $k_{p,\beta}$ and $k_{d,\beta}$, the product $\gamma_{in}\gamma_{out}$ can be made smaller than one at all times if the initial condition satisfies (\ref{smallgaininitialcondition}) since, in this case, the supremum norm of $\tilde{\theta}$ satisfies $||\tilde{\theta}||_\infty\leq\tilde{\theta}_{max}$. Therefore, the Small Gain Theorem applies and the closed loop system is ISS with respect to the UGV acceleration $\ddot{x}$. Since for a constant reference this acceleration tends to zero, asymptotic stability follows.  
\end{proof}
\end{proposition}
The previous Proposition proves that the system is asymptotically stable in absence of the saturations (\ref{actuatorsconstraints}) for all the points of equilibrium, i.e. for any $\bar{x}\in\mathbb{R}$ and $\bar{\alpha}\in[\alpha_{min},\alpha_{max}]$. In the presence of the saturations (\ref{actuatorsconstraints}), it can be proved that all the previous stability results remain valid for $\bar{\alpha}\in[\alpha_{min}+\mu,\alpha_{max}-\mu]$, where $\mu\in\mathbb{R}_0^+$ is arbitrarily small.
\begin{corollary}\label{corollaryProposition3}
Consider (\ref{UGV}) controlled by (\ref{u3}), (\ref{ftexplicit})-(\ref{maptheta}) and (\ref{u2control}). For $\bar{x} \in \mathbb{R}$ and  $\bar{\alpha} \in [\alpha_{min}+\mu, \alpha_{max}-\mu]$, the point of equilibrium $[\bar{x}, \bar{\alpha}, \bar{\beta}]^T$  is asymptotically stable in presence of the saturations (\ref{actuatorsconstraints}) where the initial condition satisfies (\ref{smallgaininitialcondition}).
\begin{proof}
In presence of the saturations (\ref{actuatorsconstraints}), it is enough to note that, since the control laws are continuous, for any point of equilibrium $\bar{\alpha}\in[\alpha_{min}+\mu,\alpha_{max}-\mu]$, there always exists a suitably small invariant set for which saturations do not occur. As a consequence, the same results of Proposition \ref{smallgainproposition} apply.
\end{proof}
\end{corollary}
Interestingly enough, it is possible to improve this control law by substituting (\ref{mapu1}) with
\begin{equation}\label{u1new}
u_1=\sigma_{0,U_{max}}\left(\dfrac{f_t}{\sin\theta}\right),
\end{equation}
where $\sigma_{0,U_{max}}$ is the positive saturation function limiting the thrust to $U_{max}$ (cf. Definition \ref{strictpositivesaturation}). The following Proposition proves that the new control law (\ref{u1new}) improves (\ref{mapu1}) and that the system is still asymptotically stable.
\begin{proposition}
Consider (\ref{UGV}) controlled by (\ref{u3}),(\ref{ftexplicit}),(\ref{maptheta}),(\ref{u2control}) and (\ref{u1new}). For $\bar{x} \in \mathbb{R}$ and $\bar{\alpha} \in [\alpha_{min}+\mu, \alpha_{max}-\mu]$, the point of equilibrium $[\bar{x}, \bar{\alpha}, \bar{\beta}]^T$  is asymptotically stable, where the initial condition satisfies (\ref{smallgaininitialcondition}). Moreover, the control law (\ref{u1new}) is equivalent to a feedforward that reduces the gain of the inner loop $\gamma_{in}$ by delivering a smaller attitude error $\tilde{\theta}_f$ to the outer loop, i.e. an attitude error that satisfies $|\tilde{\theta}_f|\leq|\tilde{\theta}|$.
\begin{proof}\label{propositionu1new}
The proof can be found in Appendix \ref{proofu1new}.
\end{proof}
\end{proposition}

In the next section, the control scheme will be improved by making use of the nonlinear Reference Governor (RG). In fact, the system can be made asymptotically stable for a larger set of initial conditions by enforcing the constraints with the RG.

\section{Constraints Enforcement}
In this section, the control law studied in the previous section will be augmented with the nonlinear RG introduced in \cite{bemporad1998reference} to avoid constraints violation. The RG can be summarized as follows. Let the desired position and angle references $[\bar{x},\bar{\alpha}]$ be given, where $\bar{\alpha}\in[\alpha_{min}+\mu,\alpha_{max}-\mu]$ and $\mu$ an arbitrary (small) positive scalar. If needed, the RG substitutes the desired set-point $[\bar{x},\bar{\alpha}]$ with a sequence of applied way-points $[\bar{\alpha},\bar{x}]_k$ which do not make the system violate the constraints. This sequence is computed online as follows. Assume that at time $t=k$, the applied reference $[\bar{x}, \bar{\alpha}]_k$, if maintained constant, would not violate the constraints. The RG computes (at fixed time intervals) the next applied reference
\begin{equation}\label{RG}
%
[\bar{x},\bar{\alpha}]_{k+1}=(1-c)[\bar{x},\bar{\alpha}]_k+c[\bar{x},\bar{\alpha}]
%
\end{equation}
by maximizing the scalar $c \in [0\,\, 1]$ under the condition that if $[\bar{x},\bar{\alpha}]_{k+1}$ is kept constant, the system would not violate constraints at any future time instant. The optimization of $c$ can be performed using bisection \cite{bemporad1998reference} and online simulations over a sufficiently long prediction horizon. The convexity of the steady-state admissible equilibria ensures that the way-point sequence converges to $[\bar{x},\bar{\alpha}]$.

\section{Simulations}
\begin{figure}[!ht]\centering{
\includegraphics[scale=0.6]{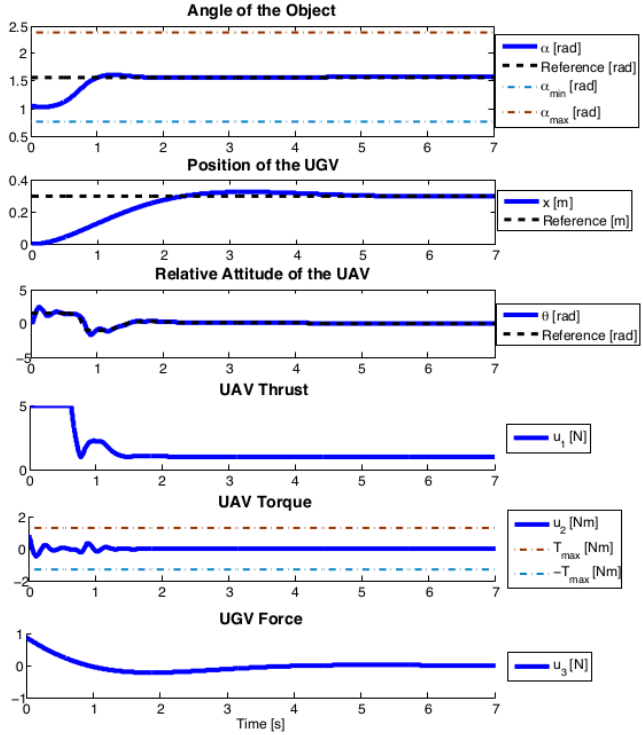}
\caption{States and inputs evolution for $\mathcal{I}_u=0.881[g.m]$ and $\bar{\alpha}=\pi/2$.}\label{firstcontroller}}
\end{figure}

Consider a UAV of mass $m_u=200\text{[g]}$ and of inertia $\mathcal{I}_u=0.881\text{[g.m$^2$]}$, cooperating with a UGV of mass $m_c=2\text{[kg]}$ to manipulate an object of mass $m_b=1\text{[kg]}$ and of inertia $\mathcal{I}_b=0.33\text{[kg.m$^2$]}$. The saturations of the actuators are $U_{max}=5\text{[N]}$, $T_{max}=1.3\text{[Nm]}$ and $F_{max}=10\text{[N]}$. The system is controlled using (\ref{u3}), (\ref{ftexplicit})-(\ref{maptheta}) and (\ref{u2control}), with $k_{p,x}=3$, $k_{d,x}=3$, $k_{p,\alpha}=20$, $k_{d,\alpha}=5$, $k_{p,\beta}=0.5$, $k_{d,\beta}=0.01$, and $\epsilon=1$. The initial condition of the system is $[x(0),\dot{x}(0),\alpha(0),\dot{\alpha}(0),\beta(0),\dot{\beta}(0)]^T=[0,0,\pi/3,0,\pi/4,0]^T$ and the desired references for the object are $\bar{\alpha}=\pi/2$ and $\bar{x}=0.3[\text{m}]$. Fig. \ref{firstcontroller} depicts the evolution of the states $[x,\alpha,\theta]^T$ and of the inputs $u_1$, $u_2$ and $u_3$. It is seen that the states are converging to the desired references and that the UAV thrust input does not violate the constraint on the thrust positiveness.

\begin{figure}[!ht]\centering{
\includegraphics[scale=0.6]{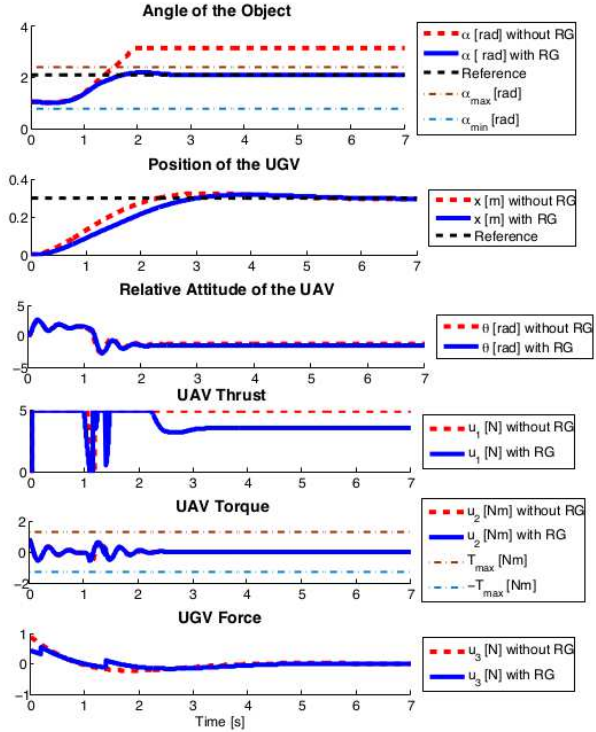}
\caption{States and inputs evolution for $\mathcal{I}_u=1.762[g.m]$ and $\bar{\alpha}=2\pi/3$ using the control law (\ref{u1new}).}\label{RGstates}}
\end{figure}

Finally, consider the evolution of the system for the desired reference $\bar{\alpha}=2\pi/3$. For the initial condition $[x(0),\dot{x}(0),\alpha(0),\dot{\alpha}(0),\beta(0),\dot{\beta}(0)]^T=[0,0,\pi/3,0,\pi/4,0]^T$, the system is unstable because the overshoot of $\alpha$ violates the constraints (see first subplot of Fig. \ref{RGstates} (red-dashed lines)). In fact, the object goes beyond the constraints and falls down to $\alpha=\pi$ since the system cannot recover anymore. This is why, to enforce the constraints and make the system asymptotically stable, the RG (\ref{RG}) is implemented with a sampling time of $t_s=0.2$[s] (see Fig. \ref{RGstates} (blue continuous line)).

\section{Conclusions}
The paper introduces a control strategy where a UAV and a UGV collaborate to manipulate an object. 
In particular, a scheme is proposed where the UGV is in charge of the position of the object and the UAV of its inclination. The stability of this  scheme is proved through Input-to-State Stability and Small Gain theorem arguments. To ensure constraints satisfaction, a nonlinear Reference Governor is added to the control scheme. Numerical simulations show the effectiveness of the proposed control strategy. Future works will aim at extending the results of this paper to the three-dimensional case.
\bibliographystyle{IEEEtran}
\bibliography{Bibliography.bib}
\appendices

\section{Proof of Lemma \ref{u1coslemma}}\label{u1coslemmaproof}
Following from the Triangular Inequality, Eq. (\ref{deltatildeequation}) is bounded by
\begin{equation}\label{triangularinequality}
|\delta_{\tilde{\theta}}|\leq|u_1\cos\bar{\theta}||\sin\tilde{\theta}|+Mg|\cos\alpha||1-\cos\tilde{\theta}|.
\end{equation}
First, note that the term $Mg|\cos\alpha||1-\cos\tilde{\theta}|$ is clearly bounded by $Mg|1-\cos\tilde{\theta}|$. For what concerns the term $|u_1\cos{\bar{\theta}}||\sin{\tilde{\theta}}|$ in (\ref{triangularinequality}), following from (\ref{mapu1}) and (\ref{maptheta}), $u_1\cos{\bar{\theta}}$ is
\begin{equation}\label{u1cos}
u_1\cos{\bar{\theta}}=\dfrac{f_t}{\tan(\sigma_{\pi/2}(\gamma\arctan(\epsilon f_t)))}.
\end{equation}
For $u_1\not\in[-U_{max},U_{max}]$, due to the saturation $\sigma_{\pi/2}$, $u_1\cos{\bar{\theta}}=0$ since $\lim_{\bar{\theta}\to\pm\pi/2}\frac{f_t}{\tan\bar{\theta}}=0$. As for $f_t$ restricted to $[-U_{max},U_{max}]$, it is easy to see that (\ref{u1cos}) is continuous as the only potential singularity admits a finite limit, which is
\begin{equation}\label{continuous1}
\lim_{f_t\to0}\dfrac{f_t}{\tan(\gamma\arctan(\epsilon f_t))}=\dfrac{1}{\gamma\epsilon}.
\end{equation}
Since $u_1\cos\bar{\theta}$ is continuous and differentiable in the closed interval $f_t\in[-U_{max},U_{max}]$, the possible extrema of (\ref{u1cos}) can be found at the boundaries $f_t=\pm U_{max}$ and at the stationary points, where $\dfrac{d}{df_t}u_1\cos{\bar{\theta}}=0$. In fact, the only point where $\dfrac{d}{df_t}u_1\cos{\bar{\theta}}=0$ is $f_t=0$. Therefore, since $u_1\cos\bar{\theta}\bigg|_{f_t=\pm U_{max}}=0$ and $u_1\cos\bar{\theta}\bigg|_{f_t=0}=\dfrac{1}{\gamma\epsilon}$, (\ref{u1cos}) reaches its maximum when $f_t=0$. In particular, since $\dfrac{1}{\gamma\epsilon}$ is strictly decreasing for $\epsilon\in\mathbb{R}_0^+$, the maximum of $\dfrac{1}{\gamma\epsilon}$ is $\lim_{\epsilon\to0}\dfrac{1}{\gamma\epsilon}=2/\pi$. Consequently, $|\delta_{\tilde{\theta}}|\leq(2/\pi|\sin\tilde{\theta}|+Mg|1-\cos\tilde{\theta}|)$ for any $\epsilon\in\mathbb{R}_0^+$ and for any $f_t\in\mathbb{R}$.

\section{Proof of Proposition \ref{propositionLyapunov}}\label{proofLyapunov}
\begin{definition}
A function $\alpha:\mathbb{R}_0^+\times\mathbb{R}_0^+\to\mathbb{R}_0^+$ is of class $\mathscr{K}_\infty$ if it is continuous, positive definite, strictly increasing, and unbounded.
\end{definition}

Define the object inclination error $\tilde{\alpha}:=\alpha-\bar{\alpha}$ and the state $x_\alpha:=[\tilde{\alpha},\dot{\alpha}]^T$. Consider
\begin{equation}\label{neweqalpha}
\ddot{\alpha}=(-k_p\tilde{\alpha}-k_d\dot{\alpha})\cos\theta+\delta,
\end{equation}
where $k_p:=k_{p,\alpha}/\mathcal{I}_0$, $k_d:=k_{d,\alpha}/\mathcal{I}_0$, and $\delta:=1/\mathcal{I}_0(\delta_{\tilde{\theta}}+d$). Define as a Lyapunov function
\begin{equation}\label{lyapunovfunction}
V=\dfrac{1}{2}x_\alpha^T\left[\begin{array}{cc}(k_p+\epsilon k_d)\cos\tilde{\theta_{max}} & \epsilon\\ 
\epsilon & 1\end{array}\right]x_\alpha,
\end{equation}
where $\epsilon\in(0,k_d\cos\tilde{\theta}_{max})$. The square matrix in (\ref{lyapunovfunction}) is clearly positive definite and, by substituting (\ref{neweqalpha}) in (\ref{lyapunovfunction}), the derivative of $V$ is
\begin{equation}
\begin{aligned}
\dot{V}=&-x_\alpha^T \left[\begin{array}{cc}\epsilon k_pc_\theta & (k_p+\epsilon k_d)(c_{\theta-{\tilde{\theta}_{max}}})\\
(k_p+\epsilon k_d)(c_{\theta-{\tilde{\theta}_{max}}}) & 2k_dc_\theta-\epsilon\end{array}\right]x_\alpha\\&+(\dot{\alpha}+\epsilon\tilde{\alpha})\delta
\end{aligned}
\end{equation}
where $c_{\theta-\tilde{\theta}_{max}}=1/2(\cos\theta-\cos\tilde{\theta}_{max})$. $\dot{V}$ can be bounded by
\begin{equation}\label{putainmerde}
\dot{V}\leq-x_\alpha Q x_\alpha + (\dot{\alpha}+\epsilon\tilde{\alpha})\delta,
\end{equation}
where $Q:=\left[\begin{array}{cc}\epsilon k_p \cos\tilde{\theta}_{max} &r\\
 r & 2k_d\cos\tilde{\theta}_{max}-\epsilon\end{array}\right]$
 and $r:=\dfrac{1}{2}(k_p+\epsilon k_d)(1-\cos\tilde{\theta}_{max})$. To ensure that $Q$ is positive definite, the inequality
\begin{equation}\label{trucdefou}
(\epsilon k_p\cos\tilde{\theta}_{max})(2 k_d\cos{\tilde{\theta}}_{max}-\epsilon) > \dfrac{1}{4}(k_p+\epsilon k_d)^2(1-\cos\tilde{\theta}_{max})^2
\end{equation}
must be imposed. To simplify the computations, let us denote $k_p=\omega^2$ and $k_d=2\xi\omega$. Substituting $\epsilon=2\xi\omega\cos\tilde{\theta}_{max}\nu$ for $\nu\in(0,1)$ in (\ref{trucdefou}), the inequality becomes
\begin{equation}\label{putain}
4\xi^2\nu^2>\dfrac{1}{4}(1+8\nu\xi^2+16\nu^2\xi^4)\dfrac{(1-\cos{\tilde{\theta}_{max}})^2}{\cos^2\tilde{\theta}_{max}}.
\end{equation}
In view of (\ref{putainmerde}) and (\ref{putain}), given $\xi$, there exists $\nu$ and $\tilde{\theta}_{max}$ such that the Lyapunov function $V$ is strictly decreasing for $\delta=0$. To prove ISS, it is sufficient to note that
\begin{equation}
\begin{aligned}
\begin{cases}
(\dot{\alpha}+\epsilon\tilde{\alpha})\delta= x_\alpha^TR\delta &\leq ||x_\alpha^T||\;||R\delta||\\
x_\alpha^T Q x_\alpha &\geq \lambda_Q||x_\alpha||^2,
\end{cases}
\end{aligned}
\end{equation}
where $$R=\left[\begin{array}{c}\epsilon \\ 1\end{array}\right]$$ and $\lambda_Q$ is the lowest eigenvalue of the positive defi􏰂nite matrix Q. As a result,
\begin{equation}
||x_\alpha||\geq||R\delta||
\end{equation}
implies $\dot{V}\leq 0$, thus proving ISS with restriction $\tilde{\theta}\in[-\tilde{\theta}_{max},\tilde{\theta}_{max}]$ with respect to $\delta$. Since the system is ISS with restriction $\tilde{\theta}\in[-\tilde{\theta}_{max},\tilde{\theta}_{max}]$ with respect to $\delta$, it follows ISS with restriction $\tilde{\theta}\in[-\tilde{\theta}_{max},\tilde{\theta}_{max}]$ with respect to $\delta_{\tilde{\theta}}$ and $d$.

To prove that the system is ISS with restriction $\tilde{\theta}\in[-\tilde{\theta}_{max},\tilde{\theta}_{max}]$ with respect to $\tilde{\theta}$, it is enough to find a gain $\Gamma$ between $\delta_{\tilde{\theta}}$ and $\tilde{\theta}$ that is a function of class $\mathscr{K}_\infty$. In fact, following from Lemma \ref{u1coslemma}, the disturbance $\delta_{\tilde{\theta}}$ is bounded by
\begin{equation}
|\delta_{\tilde{\theta}}|\leq2/\pi|\sin\tilde{\theta}|+Mg|1-\cos\tilde{\theta}|.
\end{equation}
The second member of the inequality is bounded by
\begin{equation}
2/\pi|\sin\tilde{\theta}|+Mg|1-\cos\tilde{\theta}|\leq\Gamma(\tilde{\theta}),
\end{equation}
where $\Gamma(\tilde{\theta}):=2/\pi|\tilde{\theta}|+Mg\tilde{\theta}^2$ is a function of class $\mathscr{K}_\infty$. As a consequence, since the gain between $\tilde{\theta}$ and $\delta_{\tilde{\theta}}$ is a function of class $\mathscr{K}_\infty$ and the system is ISS with restriction $\tilde{\theta}\in[-\tilde{\theta}_{max},\tilde{\theta}_{max}]$ with respect to $\delta_{\tilde{\theta}}$, the system is also ISS with restriction $\tilde{\theta}\in[-\tilde{\theta}_{max},\tilde{\theta}_{max}]$ with respect to $\tilde{\beta}$ (cf. Eq. (\ref{thetaequalbeta})).

It remains to prove that the gain between $\tilde{\beta}$ and $\dot{\bar{\beta}}$ exists and is finite. The derivative of $\bar{\beta}$ is
\begin{equation}\label{derivbeta}
\dot{\bar{\beta}}=\dot{\bar{\theta}}+\dot{\alpha}.
\end{equation}
Since System (\ref{u1experrorcanonic}) is ISS with restriction $\tilde{\theta}\in[-\tilde{\theta}_{max},\tilde{\theta}_{max}]$ with respect to $\tilde{\beta}$, it follows that $|\dot{\alpha}|\leq\xi(\tilde{\beta})$ and $|\alpha|\leq\xi(\tilde{\beta})$, where $\xi$ is a function of class $\mathscr{K}_\infty$. Consequently, the acceleration of the object in (\ref{u1experrorcanonic}) is also bounded and thus
\begin{equation}\label{alphabounded}
|\ddot{\alpha}|\leq\Delta(\tilde{\beta}),
\end{equation}
where $\Delta$ is a function of class $\mathscr{K}_\infty$. The derivative of (\ref{maptheta}) is
\begin{equation*}
\dot{\bar{\theta}}=\dfrac{\gamma\epsilon\dot{f_t}}{1+(\epsilon f_t)^2}.
\end{equation*}
Since $(\epsilon f_t)^2\geq0$, the inequality
\begin{equation*}
|\dot{\bar{\theta}}|\leq\gamma\epsilon|\dot{f_t}|
\end{equation*}
holds true. Expressing the time derivative of $f_t$, it can be said that
\begin{eqnarray}\label{thetabounded}
\begin{aligned}
|\dot{\bar{\theta}}|&\leq\gamma\epsilon|-k_{p,\alpha}\dot{\alpha}-k_{d,\alpha}\ddot{\alpha}-M\sin{\alpha}\dot{\alpha}g|\\
&\leq\gamma\epsilon(|k_{p,\alpha}\dot{\alpha}+k_{d,\alpha}\ddot{\alpha}|+|Mg\dot{\alpha}|)\\
&\leq\gamma\epsilon(k_{p,\alpha}+k_{d,\alpha}+Mg)\zeta(\tilde{\beta}),
\end{aligned}
\end{eqnarray}
where $\zeta(\tilde{\beta}):=\max(\xi(\tilde{\beta}),\Delta(\tilde{\beta}))$. Therefore, $\dot{\bar{\beta}}$ is bounded with respect to $\tilde{\beta}$ in view of (\ref{derivbeta}), (\ref{alphabounded}) and (\ref{thetabounded}). Consequently, there exists a finite asymptotic gain $\gamma_{out}$ between the disturbance $\tilde{\beta}$ and the output $\dot{\bar{\beta}}$.

\section{Proof of Proposition \ref{UAVAttitudeProposition}}\label{proofUAVAttitude}

Since (\ref{attitudedynamics}) is linear and asymptotically stable for any $k_{p,\beta}>0$ and $k_{d,\beta}>0,$ it is also ISS and the asymptotic gain $\gamma_{in}$ is the $l_1$ norm between $\dot{\bar{\beta}}$ and $\tilde{\beta}.$  To prove that this gain can be made arbitrarily small, consider the parameter choice $k_{p,\beta}/\mathcal{I}_u=\omega^2$ and $k_{d,\beta}/\mathcal{I}_u=2\omega$ with $\omega\in\mathbb{R}_0^+.$ It results that the impulsive response between $\dot{\bar{\beta}}$ and  $\tilde{\beta}$ is 
\[
\begin{array}{lll}
w(t) & = &\left[\begin{array}{cc} 1 & 0 \end{array}\right]exp\left(\left[\begin{array}{cc} 0 & 1  \\-\omega^2 & -2\omega \end{array}\right]t\right) \left[\begin{array}{c} 1 \\ 0 \end{array}\right] \\ 
&=& -(1+\omega t)e^{-\omega t}.
\end{array}
\]
By the definition of the $l_1$ norm, the gain $\gamma_{in}$ is 
\[
\gamma_{in}= \int_{0}^\infty | w(t) | dt =  \frac{1}{\omega},
\]
which can be made arbitrarily small for a suitably large $\omega$ and therefore for suitably large $k_{p,\beta}/\mathcal{I}_u$ and $k_{d,\beta}/\mathcal{I}_u$. Finally, remark that $\gamma_{in}$ also depends on the UAV inertia $\mathcal{I}_u$.

\section{Proof of Proposition \ref{propositionu1new}}\label{proofu1new}
To prove $|\tilde{\theta}_f|\leq|\tilde{\theta}|$, first let $f_t$ denote the reference for the tangential force that is requested by the system. Let $f_{old}$ and $f_{new}$ be the actual forces delivered to the object using the control laws (\ref{mapu1}) and (\ref{u1new}), respectively, which are
\begin{equation}\label{fnewfold}
\begin{aligned}
\begin{cases}
f_{old}&:=\dfrac{f_t\sin\theta}{\sin\bar{\theta}}\\&=\dfrac{f_t\sin(\bar{\theta}+\tilde{\theta})}{\sin{\bar{\theta}}}\\
f_{new}&:=\sigma_{0,U_{max}}\left(\dfrac{f_t}{\sin\theta}\right)\sin\theta.
\end{cases}
\end{aligned}
\end{equation}
The control law (\ref{u1new}) can be seen as a feedforward block on the control law (\ref{mapu1}), which generates a fictitious attitude error $\tilde{\theta}_f$ instead of $\tilde{\theta}$, where $\tilde{\theta}_f$ is such that
\begin{equation}\label{thetaf}
\dfrac{f_t\sin(\bar{\theta}+\tilde{\theta}_f)}{\sin\bar{\theta}}=\sigma_{0,U_{max}}\left(\dfrac{f_t}{\sin\theta}\right)\sin\theta.
\end{equation}
Under the assumption that $f_t\leq U_{max}$, there are three cases:
\begin{enumerate}
\item If $\text{sign}(\sin(\bar{\theta}+\tilde{\theta}))=-\text{sign}(\sin\bar{\theta})$, then $f_{new}=0$, which corresponds to the case where $\tilde{\theta}_f=-\bar{\theta}$ using the previous mapping (\ref{mapu1}) (see Eq. (\ref{fnewfold}) and (\ref{thetaf})). Note that $$\text{sign}(\sin\theta)=-\text{sign}(\sin\bar{\theta})$$ implies that $|\tilde{\theta}|\geq|\bar{\theta}|$ in the first law. As a consequence, $|\tilde{\theta}_f|\leq|\tilde{\theta}|$.
\item If $\dfrac{f_t}{\sin\theta}\geq U_{max}$, then $f_{new}=U_{max}\sin\theta$. It is worth to remark that
\begin{equation}\label{casetwoproof}
\begin{aligned}
\dfrac{f_t\sin(\bar{\theta}+\tilde{\theta})}{\sin\bar{\theta}}&\leq \dfrac{f_t\sin(\bar{\theta}+\tilde{\theta}_f)}{\sin\bar{\theta}}\leq f_t.
\end{aligned}
\end{equation}
The inequalities (\ref{casetwoproof}) are equivalent to
\begin{equation}
\begin{aligned}
\left|\dfrac{\sin(\bar{\theta}+\tilde{\theta})}{\sin\bar{\theta}}\right|&\leq\left|\dfrac{\sin(\bar{\theta}+\tilde{\theta}_f)}{\sin\bar{\theta}}\right|\leq1.
\end{aligned}
\end{equation}
As a consequence, $0\leq|\tilde{\theta}_f|\leq|\tilde{\theta}|$.
\item In all the other cases, no saturation occurs and $f_{new}=f_t$. In view of (\ref{fnewfold}) and (\ref{thetaf}), this case is equivalent to the first control law (\ref{mapu1}) where $\tilde{\theta}_f=0$.
\end{enumerate}
As a result, since $|\tilde{\theta}_f|\leq|\tilde{\theta}|$, the gain between $\dot{\bar{\beta}}$ and $\tilde{\theta}_f$ is even smaller and all the stability results of Proposition \ref{smallgainproposition} and Corollary \ref{corollaryProposition3} apply.

\end{document}